\documentclass[5p,singleauthorgroup]{elsarticle}

\usepackage{lineno,hyperref}
\usepackage{epstopdf}
\modulolinenumbers[5]

\journal{Astroparticle Physics}

\begin{document}

\begin{frontmatter}

\title{VERITAS Detection of LS 5039 and HESS J1825-137}

\author[1]{A.~U.~Abeysekara}
\author[2]{W.~Benbow}
\author[3]{R.~Bird}
\author[4,5]{R.~Brose}
\author[6]{J.~L.~Christiansen}
\author[7]{A.~J.~Chromey}
\author[8,9]{W.~Cui}
\author[2]{M.~K.~Daniel}
\author[10]{A.~Falcone}
\author[11]{L.~Fortson}
\author[12]{D.~Hanna}
\author[5]{T.~Hassan}
\author[13]{O.~Hervet}
\author[14]{J.~Holder}
\author[2]{G.~Hughes}
\author[15]{T.~B.~Humensky}
\author[16]{P.~Kaaret}
\author[1]{P.~Kar}
\author[5]{N.~Kelley-Hoskins}
\author[17]{M.~Kertzman}
\author[1]{D.~Kieda}
\author[5]{M.~Krause}
\author[18]{M.~J.~Lang}
\author[5]{G.~Maier}
\author[18]{P.~Moriarty}
\author[19]{D.~Nieto}
\author[5]{M.~Nievas-Rosillo}
\author[3]{R.~A.~Ong}
\author[20]{D.~Pandel}
\author[4,5]{M.~Pohl}
\author[5]{R.~R.~Prado}
\author[5]{E.~Pueschel}
\author[21]{J.~Quinn}
\author[12]{K.~Ragan}
\author[22]{P.~T.~Reynolds}
\author[14]{G.~T.~Richards}
\author[2]{E.~Roache}
\author[5]{I.~Sadeh}
\author[23]{M.~Santander}
\author[8]{G.~H.~Sembroski}
\author[7]{A.~Weinstein}
\author[16]{P.~Wilcox}
\author[13]{D.~A.~Williams}
\author[14]{T.~J~Williamson}

\address[1]{Department of Physics and Astronomy, University of Utah, Salt Lake City, UT 84112, USA}
\address[2]{Center for Astrophysics $|$ Harvard \& Smithsonian, Cambridge, MA 02138, USA}
\address[3]{Department of Physics and Astronomy, University of California, Los Angeles, CA 90095, USA}
\address[4]{Institute of Physics and Astronomy, University of Potsdam, 14476 Potsdam-Golm, Germany}
\address[5]{DESY, Platanenallee 6, 15738 Zeuthen, Germany}
\address[6]{Physics Department, California Polytechnic State University, San Luis Obispo, CA 94307, USA}
\address[7]{Department of Physics and Astronomy, Iowa State University, Ames, IA 50011, USA}
\address[8]{Department of Physics and Astronomy, Purdue University, West Lafayette, IN 47907, USA} \address[9]{Department of Physics and Center for Astrophysics, Tsinghua University, Beijing 100084, China.}
\address[10]{Department of Astronomy and Astrophysics, 525 Davey Lab, Pennsylvania State University, University Park, PA 16802, USA}
\address[11]{School of Physics and Astronomy, University of Minnesota, Minneapolis, MN 55455, USA}
\address[12]{Physics Department, McGill University, Montreal, QC H3A 2T8, Canada}
\address[13]{Santa Cruz Institute for Particle Physics and Department of Physics, University of California, Santa Cruz, CA 95064, USA}
\address[14]{Department of Physics and Astronomy and the Bartol Research Institute, University of Delaware, Newark, DE 19716, USA}
\address[15]{Physics Department, Columbia University, New York, NY 10027, USA}
\address[16]{Department of Physics and Astronomy, University of Iowa, Van Allen Hall, Iowa City, IA 52242, USA}
\address[17]{Department of Physics and Astronomy, DePauw University, Greencastle, IN 46135-0037, USA}
\address[18]{School of Physics, National University of Ireland Galway, University Road, Galway, Ireland}
\address[19]{Institute of Particle and Cosmos Physics, Universidad Complutense de Madrid, 28040 Madrid, Spain}
\address[20]{Department of Physics, Grand Valley State University, Allendale, MI 49401, USA}
\address[21]{School of Physics, University College Dublin, Belfield, Dublin 4, Ireland}
\address[22]{Department of Physical Sciences, Cork Institute of Technology, Bishopstown, Cork, Ireland}
\address[23]{Department of Physics and Astronomy, University of Alabama, Tuscaloosa, AL 35487, USA}





\begin{abstract}
With 8 hours of observations, VERITAS confirms the detection of two very high energy gamma-ray sources. The gamma-ray binary LS 5039 is detected with a statistical significance of $8.8\sigma$. The measured flux above 1~TeV is $(2.5 \pm 0.4) \times 10^{-12} \rm \, cm^{-2} \, s^{-1}$ near inferior conjunction and $(7.8 \pm 2.8) \times 10^{-13} \rm \, cm^{-2} \, s^{-1}$ near superior conjunction. The pulsar wind nebula HESS J1825-137 is detected with a statistical significance of $6.7\sigma$ and a measured flux above 1~TeV of $(3.9 \pm 0.8) \times 10^{-12} \rm \, cm^{-2} \, s^{-1}$.
\end{abstract}

\begin{keyword}
binaries: close; gamma rays: general; ISM: individual objects: HESS J1825-137; stars: individual: LS 5039
\end{keyword}

\end{frontmatter}


\section{Introduction}\label{sec:intro} 
We report observations of two known, bright, very-high-energy (VHE; E $>$ 50 GeV) gamma-ray emitting objects: the gamma-ray binary LS 5039 and the very extended pulsar wind nebula (PWN) HESS J1825-137. These two objects are separated by about $1^\circ$ on the sky and can both be observed in a single field of view with the Very Energetic Radiation Imaging Telescope Array System (VERITAS)\footnote{http://veritas.sao.arizona.edu}.

The observations are described in Section \ref{sec:obs}. Background, spectra, timing, and morphological studies are described for each source in Section \ref{sec:results}.

\section{Observations}\label{sec:obs}
VERITAS is a four-telescope array exploiting the imaging atmospheric Cherenkov technique to detect VHE gamma rays and is most sensitive to energies between 0.085--30 TeV. The array is located at the Fred Lawrence Whipple Observatory (FLWO) in southern Arizona ($\rm 31^\circ 40' N, 110^\circ 57' W,$  1.3 km above sea level). For a performance summary, see \cite{2015ICRC...34..771P}.

\begin{figure}[h!]
\begin{center}
\includegraphics[width=\columnwidth,angle=0]{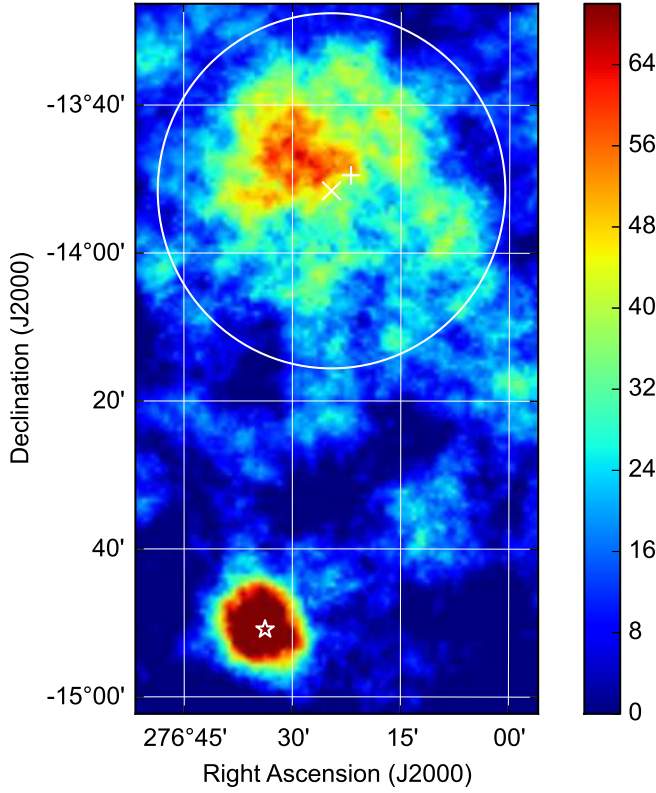}
\caption{Excess counts skymap of the region containing LS 5039 and HESS J1825-137. LS 5039 is marked with a star and HESS J1825-137 is enclosed within a $0.4^\circ$ radius circle. A cross and x respectively mark the VERITAS and HESS \citep{2019A&A...621A.116H} centroids of the PWN.}
\label{skymap}
\end{center}
\end{figure}

VERITAS observed the field in late spring of 2013 and 2014 with an average zenith angle of $48^\circ$. Standard analysis was performed \citep{2008ICRC....3.1325D} with background extracted from regions near the sources excluding a $0.35^\circ$ radius region around LS 5039 and a $0.75^\circ$ radius region around HESS J1825-137. With 7.9~hours of usable observations after quality cuts, we confirm the HESS detections of these sources \citep{2005Sci...309..746A,2005A&A...442L..25A}. Figure \ref{skymap} shows the VERITAS skymap made with an integration region radius of $\theta = 0.089^\circ$. The average low-energy threshold for VERITAS observations is 420 GeV due to the relatively large zenith angle of the observations.

\section{Results}\label{sec:results}
\subsection{LS 5039}\label{sec:ls5039}
LS~5039 was identified as a high mass X-ray binary in the ROSAT Galactic plane survey \citep{1997A&A...323..853M} and radio emission was detected using the VLA \citep{1998A&A...338L..71M}. It was also detected as a gamma-ray binary with EGRET \citep{2000Sci...288.2340P} and at VHE with HESS \citep{2005Sci...309..746A}. Periodic emission is also detected by \textit{Fermi}-LAT above 300 MeV \citep{2017MNRAS.471.3036P}. The companion star is spectral type O6.5V \citep{2001A&A...376..476C}, while the mass and nature of the compact object is undetermined \citep{2018MNRAS.474.4756Y}. However, some evidence exists for a neutron star being present in the system. Orbital modulation in TeV gamma-rays peaks when the star and compact object are aligned relative to the observer which can be explained by the outward flow of relativistic particles from a PWN boosting emission from the star \citep{2014ApJ...790...18T}. 

\begin{figure}[h!]
\begin{center}
\includegraphics[width=\columnwidth,angle=0]{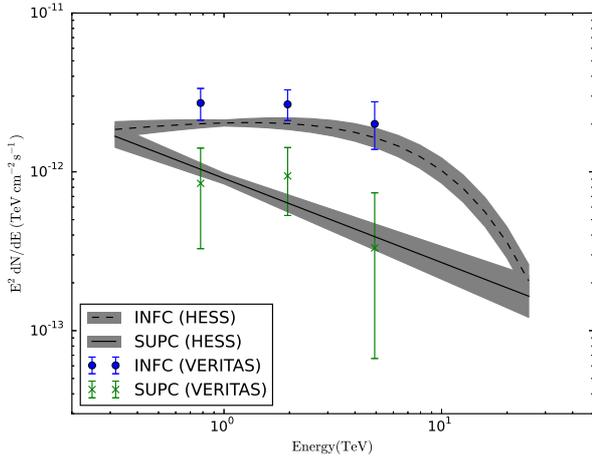}
\caption{Spectra of LS 5039 separated by orbital phase. The filled circles and dashed line are at inferior conjunction. The x's and solid lines are at superior conjunction. Points are from this analysis and the fit spectra with shaded errors are from \cite{2006A&A...460..743A}.  }
\label{ls5039sed}
\end{center}
\end{figure}

We detect VER J1826-148, the VHE source associated with LS~5039, at a significance of 8.8 standard deviations ($\sigma$) \citep{1983ApJ...272..317L} with 101 excess counts within a region of radius $\theta = 0.089^\circ$ centered at the known position from HESS $(\alpha,\delta) = (276.563^\circ ,-14.848^\circ )$ (J2000) \citep{2006A&A...460..743A}. Each observation was a single $\sim 20-30$ minute exposure, and consecutive exposures were separated by 1-2 nights. This provided approximately daily coverage of the 3.9 day orbital period \citep{2006A&A...460..743A} for 8 days in 2013 and 12 days in 2014. Spectra separated into two orbital phase intervals, see Fig.~\ref{ls5039sed}, agree within the uncertainties of those measured with HESS \citep{2006A&A...460..743A}. The flux above 1~TeV is $\rm (2.5 \pm 0.4_{stat} \pm 0.5_{sys}) \times 10^{-12} \rm \, cm^{-2} \, s^{-1}$ near inferior conjunction (orbital phase $0.45< \phi \leq 0.9$, 232 minutes exposure) and $\rm (7.8 \pm 2.8_{stat} \pm 1.6_{sys}) \times 10^{-13} \rm \, cm^{-2} \, s^{-1}$ near superior conjunction ($0.0 \leq \phi \leq 0.45$ and $0.9 < \phi \leq 1.0$, 230 minutes). This difference in flux demonstrates that the previously observed orbital flux variability \citep{2005Sci...309..746A} is detected by VERITAS as well. Fitting a power law to the spectrum for each orbital phase interval gives a photon index near inferior conjunction of $\rm \Gamma=2.1 \pm 0.2_{stat} \pm 0.2_{sys}$ with $\chi^2 /\rm DoF =0.19/1$ and $\rm \Gamma=2.4 \pm 0.5_{stat} \pm 0.2_{sys}$ with $\chi^2 /\rm DoF =0.57/1$ near superior conjunction. The VERITAS flux points match the best fitted spectral models from HESS with $\chi^2 /\rm DoF =1.60/3$ and $\chi^2 /\rm DoF =0.55/3$ for the phase intervals near inferior and superior conjunction, respectively, and support the PWN interpretation \citep{2014ApJ...790...18T}.

\subsection{HESS J1825-137}\label{sec:j1825}
The first indication of a PWN in this region was due to ROSAT observations of an extended ($\sim 5'$) X-ray nebula associated with the young (21~kyr) and energetic ($\dot{E} = 2.8\ \times 10^{36} \rm \, ergs \, s^{-1}$) pulsar PSR B1823-13 \citep{1996ApJ...466..938F}. HESS J1825-137 has a radius at least 5 times larger in VHE than X-ray, and the larger extent is consistent with the longer lifetime of TeV emitting electrons \citep{2005A&A...442L..25A}. The large angular size of HESS J1825-137 makes it ideal for resolved VHE study of PWN to better understand the particle transport within\citep{2019A&A...621A.116H}.

Using the same observations and analysis described for LS~5039, VERITAS detected HESS J1825-137, at the center of the PWN reported by HESS $(\alpha,\ \delta) = (276.421^\circ$, $-13.839^\circ )$ (J2000) \citep{2019A&A...621A.116H} with a significance of $5.3\sigma$; the peak significance of $6.7\sigma$ within a $0.4^\circ$ radius region.  Fitting a spectrum extracted from this region to a power law model gave a normalization at 1~TeV of $\rm A = (5.0 \pm 0.7_{stat} \pm 1.0_{sys}) \times 10^{-12} \rm \, cm^{-2} \, s^{-1} \, TeV^{-1}$, a photon index $\Gamma=2.28 \pm 0.15_{stat} \pm 0.2_{sys}$, and $\chi^2 /\rm DoF = 2.27/1$. We found the PWN (named VER J1825-138) to be centered at $(\alpha,\ \delta) = \rm (276.37^\circ \pm 0.02^\circ_{stat} \pm 0.01^\circ_{sys}, -13.83^\circ \pm 0.02^\circ_{stat} \pm 0.01^\circ_{sys})$ (J2000) using a weighted centroid. The region of interest, HESS centroid and VERITAS centroid are marked in Fig. \ref{skymap}.  Summing in azimuth around the PWN, the radial profile is defined by a Gaussian with $\sigma = 0.27^\circ$ (1$\sigma$ confidence interval of $\ 0.22^\circ - 0.28^\circ$). 


HESS reports a power-law fit from a $0.4^\circ$ region of $\rm A = (6.81 \pm 0.07_{stat} \pm 0.2_{sys}) \times 10^{-12} \rm \, cm^{-2} \, s^{-1} \, TeV^{-1}$ with a photon index $\Gamma=2.28 \pm 0.01_{stat} \pm 0.02_{sys}$  \citep{2019A&A...621A.116H}. The VERITAS flux normalization is consistent with that found by HESS. Similarly, the position of the PWN measured with VERITAS is within $0.04^\circ$ of that detected by HESS. The radial extent of the nebula measured by VERITAS is consistent with the original HESS observations of $\sigma = 0.24^\circ \pm 0.02^\circ$ with a ~52hr exposure \citep{2006A&A...460..365A}, but the exceptionally deep observations from HESS \citep{2019A&A...621A.116H} yields a different, larger, energy dependent extent. 

\section{Conclusion}
VERITAS has confirmed the TeV gamma-ray detections of LS 5039 and HESS J1825-137 and analysis agrees with previous findings about the energetics and morphology. 

\section{Acknowledgements}

This research is supported by grants from the U.S. Department of Energy Office of Science, the U.S. National Science Foundation and the Smithsonian Institution, and by NSERC in Canada. This research used resources provided by the Open Science Grid, which is supported by the National Science Foundation and the U.S. Department of Energy's Office of Science, and resources of the National Energy Research Scientific Computing Center (NERSC), a U.S. Department of Energy Office of Science User Facility operated under Contract No. DE-AC02-05CH11231. We acknowledge the excellent work of the technical support staff at the Fred Lawrence Whipple Observatory and at the collaborating institutions in the construction and operation of the instrument.


\bibliography{mybib}

\end{document}